\documentclass[11pt]{iopart}
\usepackage{iopams}
\usepackage{amsgen,amsbsy}
\usepackage{amsfonts,mathrsfs}

\usepackage{setstack}
\usepackage{graphicx}
\usepackage{caption}
\usepackage{xcolor}

\usepackage{enumerate}

\newcommand {\be} {\begin{equation}}
\newcommand {\ee} {\end{equation}}
\newcommand {\bea} {\begin{eqnarray}}
\newcommand {\eea} {\end{eqnarray}}
\newcommand{\subs}[1]{_{\! \! \mbox{\tiny{ #1}}}}

\begin{document}

\title{
Insights from Melvin-Kerr-Newman spacetimes}
\author[Booth]{I.~Booth$^1$, M.~Hunt$^2$, A.~Palomo-Lozano$^{1,3}$, H.~Kunduri$^1$ }
\address{$^1$Dept. of Mathematics and Statistics, Memorial University, NL, Canada}
\address{$^2$ Dept. of Physics and Physical Oceanography, Memorial University, NL, Canada}
%\address[HK]{Department of Mathematics and Statistics, Memorial University of Newfoundland}
\address{$^3$ Instituto de Ciencias F\'isicas y Matem\'aticas, Universidad Austral de Chile, Valdivia, Chile}
%\phantom{(APL)\hspace{4ex}}Department of Mathematics and Statistics, Memorial University of Newfoundland}
%
\ead{ibooth@mun.ca, mbh825@mun.ca, alberto.palomo@uach.cl, hkkunduri@mun.ca}

\begin{abstract}
We examine several aspects of black hole horizon physics using the Melvin-Kerr-Newman (MKN) family of spacetimes. 
Roughly speaking these are black holes immersed in a distorting background magnetic field and unlike the standard Kerr-Newman 
(KN) family they are not asymptotically flat. As exact solutions with horizons that can be highly distorted relative 
to KN, they provide a good testbed for ideas about and theorems constraining black hole horizons. 

We explicitly show that MKN horizons with fixed magnetic field parameter
may be uniquely specified by their area, charge and angular momentum and that the charge and angular momentum are bound by 
horizon area in the same way as for KN. As expected, extremal MKN horizons are geometrically isomorphic to 
extremal KN horizons and the geometric distortion of near-extremal horizons is constrained by their proximity to extremality. At the 
other extreme, Melvin-Schwarzschild (MS) solutions may be infinitely distorted, however for intermediate cases any non-zero charge or angular momentum restricts
distortions to be finite. These properties are in agreement with known theorems but are seen to be satisfied in interesting and non-trivial ways. 
\end{abstract}

\section{Introduction}
\label{Sec:introduction}

The basic properties of Kerr-Newman (KN) spacetimes are well-known (see for example \cite{Wald} or 
\cite{poisson2004relativist}). There is a black hole region from which no signal can be sent to infinity and the 
boundary of that region is the event horizon.  Geometrically the horizon is a non-expanding null surface
and is a Killing horizon. % (where a time-translation Killing vector field becomes null). 
Inside
the horizon are trapped surfaces and a gravitational singularity.   

Physically these black holes have well-defined notions of total mass-energy $M$, charge $Q$ and angular momentum $J$. 
Those quantities are related via the Smarr law:
\be
M = \frac{\kappa}{4 \pi G} \mathcal{A} + \Omega J + \Phi Q \, . \label{IntroSmarr}
\ee
where $\mathcal{A}$ is the area of a cross-section of the horizon and $\kappa$, $\Omega$ and $\Phi$ are 
respectively its surface gravity, angular velocity and Coulomb potential.  Variations of these quantities through the 
set of all possible KN solutions gives rise to the first law of black hole mechanics:
\be
d M = \frac{\kappa}{8 \pi G} \mathrm{d} \mathcal{A} + \Omega \mathrm{d} J + \Phi \mathrm{d} Q \label{IntroFirst}
\ee

The celebrated black hole uniqueness theorems (\cite{Chrusciel:2012jk} for a review) mean that all of these properties 
are more than just peculiarities of a particular set of solutions. Those theorems tell us that, in four dimensions, 
KN black holes are the only asymptotically flat, stationary and axisymmetric Einstein-Maxwell black holes. From the 
perspective of a black hole and its characteristic time-scale, most astrophysical processes are nearly stationary and 
nearly asymptotically flat. Thus most astrophysical black holes are (perturbed) KN black holes. 

However, most is not all. Some extreme astrophysical process, such as black hole formation or the final stages of a 
black hole merger are certainly not (perturbatively) KN: they are not nearly stationary and may also contain extra
matter fields. More generally, alternative theories of gravity contain non-KN black holes. 
Thus there are good physical reasons to study other, non-KN black holes. From a mathematical perspective it is also 
of interest to study alternative or distorted black holes in order to sharpen our understanding of mathematical
definitions of black holes and  better appreciate the degree to which well-known properties are (or are not) peculiar 
to KN solutions. 

Perhaps the easiest way to obtain non-KN black holes is to violate asymptotic flatness. 
The best-known solutions of this type are  the (vacuum) Weyl-distorted Schwarzschild spacetimes \cite{Geroch:1982bv}.
Like standard Schwarzschild, these  solutions contain Killing horizons, trapped surfaces 
inside those horizons and interior singularities. However in other ways they differ significantly. The horizons are no longer 
spherically symmetric as the distortions force them to become oblate or prolate.  
The singularity is similarly distorted \cite{Frolov:2007xi}.  While there are always trapped surfaces  close
to the singularity, for sufficiently large distortions there are no longer trapped surfaces ``just inside'' the 
horizon\cite{Pilkington:2011aj}.

The distortions in these spacetimes are induced by distortions of the asymptotic structure and by similar methods one can also 
distort both Reissner-Nordstr\"{o}m\cite{Fairhurst:2000xh,Abdolrahimi:2009db} and KN\cite{Breton:1998sr,Shoom:2015rda} solutions. 
Most generally, stationary axisymmetric electrovac spacetimes may be written as Ernst solutions\cite{Griffiths:2009dfa} and there are many solution 
generating techniques that take existing solutions and generate new ones\cite{Stephani:2003tm}.  In this paper we focus on a class of solutions generated by Harrison
transforms of KN spacetimes. These Melvin-Kerr-Newman (MKN) spacetimes describe a black hole immersed 
in a background magnetic field (see, for example, \cite{Melvin:1964xx,ernst:black, ernst:kerr,Hiscock:1980zf,Bicak:2000me,AlievGaltsov,
Gibbons:2013dna,Gibbons:2013yq}). Though quite complicated  algebraically they are conceptually relatively easy to work with as the degree of distortion is 
parameterized by a single parameter $B$ which is associated with the distorting magnetic field. MKN solutions are not asymptotically flat and 
the strength of the electromagnetic field actually grows as one moves away from the black hole. 

When interpreting these solutions,  the loss of asymptotic flatness results in other complications beyond the exact definition of a black hole. 
While surface area $A$, charge $Q$ and angular momentum $J$ remain well-defined for any axisymmetric horizon
%(as will be reviewed in Section \ref{Sec:PhysProp} they may be calculated locally on the horizon) 
the other quantities appearing in the first law become problematic. The standard definitions of surface gravity $\kappa$ and angular velocity $\Omega$ depend on 
the normalization of a  global ``time'' and its associated Killing vector field at infinity. The preferred gauge for the Coulomb potential $\Phi$ also references infinity.  
Without asymptotic flatness it is no longer clear how to do any of this. 

Similarly, the standard ADM mass is 
only defined for asymptotically flat\cite{Wald} or asymptotically anti-deSitter\cite{Abbott:1981ff,Ashtekar:1999jx} 
spacetimes. There is no universally agreed method for defining the mass of spacetimes with exotic asymptotics and  it is not even obvious that 
the mass of such a spacetime even should be well-defined.
For an asymptotically flat spacetime the ADM mass is equivalent to the Newtonian mass as measured in the 
weak field zone (far from the source the gravitational field is essentially Newtonian) however for other asymptotics  there is no such region. 

%The asymptotic structure is also key to defining the surface gravity, Coulomb potential and angular velocity of a black hole. 
%If $\xi_o$ is the Killing vector field that becomes null on a Killing horizon, then the surface gravity $\kappa$ 
%and Coulomb potential $\Phi$ are:
%\be
%\kappa^2 = -\frac{1}{2} \left( \nabla_a \xi^o_b \right) \left( \nabla^a \xi_o^b \right) \; \; \mbox{and} \; \; 
%\Phi = \xi_o^a A_a \, ,
%\ee
%where $A_a$ is the electromagnetic vector potential (for notational clarity the subscript ``$o$'' is turned into a superscript for the one-form version). 
%But there is a rescaling freedom in $\xi_o$ and gauge freedom in 
%$A_a$. In the special case of a static asymptotically flat spacetime (things are slightly more complicated for the stationary case), 
%the rescaling freedom is removed by normalizing $\xi_o$ so that it is unit length at infinity  while the gauge freedom
%is removed by requiring that $A_a$ vanish at infinity. However it is not clear what the corresponding behaviours should be for exotic
%asymptotics where there may be either no or many timelike Killing vector fields at infinity and where the electromagnetic field 
%itself may diverge.   For non-asymptotically flat spacetimes that are
%stationary (and rotating) rather than static the problems are compounded as the angular velocity 
%$\Omega$ joins the list of ambiguous quantities: it is usually defined by angular component of $\xi_o$ relative to the Killing vector field that is timelike
%and unit length at infinity. 

With all of these uncertainties the status of the Smarr relation (\ref{IntroSmarr}) and first law (\ref{IntroFirst}) is unclear:
many of the quantities may no longer be well-defined. While there are proposals to deal with the uncertainties (such as the isolated horizon formalism 
\cite{Ashtekar:2000hw,Ashtekar:2001is} or a recent MKN-specific proposal \cite{Gibbons:2013dna})
the problem is not fully resolved. 
%such as the isolated horizon formalism 
%\cite{Ashtekar:2000hw,Ashtekar:2001is} or a recent MKN-specific proposal to by 
%Gibbons, Pang and Pope\cite{Gibbons:2013dna} propose ways to resolve these uncertainties (and we will consider these in 
%Section \ref{MassFL}), considerable uncertainty remains. 
That said, not all structure is lost:  \emph{horizon} uniqueness and constraint theorems remain. Most of these have been proved for 
marginally outer trapped surfaces (MOTS) which provide the standard quasilocal characterization of a black hole boundary: examples of these include 
apparent horizons as well as instantaneous slices of Killing horizons and isolated horizons. Then with the extra assumption that the MOTS
be stable (essentially there are trapped surfaces ``just inside'' the horizon) and axisymmetric it has been shown \cite{Clement:2011np,Clement:2012vb} 
that there is a universal bound that is identical in form to that for KN horizons:
\be
Q^4 + 4 J^2 \leq R^4 \, , \label{B2} 
\ee
where $R=\sqrt{\mathcal{A}/4\pi}$ is the areal radius of a cross-section of the horizon and $J$ is the total angular momentum.  This bound will then also hold for any axisymmetric distorted horizon that is stable and marginally outer trapped (the MKN horizons satisfy this condition). 

It has also been known for quite a while that the KN family of extremal Killing horizons are the unique extremal horizons in 
four-dimensional electrovac spacetimes \cite{Hajicek:1974oua, Lewandowski:2002ua, Kunduri:2007vf}. That is, if (\ref{B2}) is saturated then 
the intrinsic geometry along with certain components of the extrinsic curvature of the horizon and electromagnetic field at the horizon are identical to those 
of a member of the KN family\footnote{Examples of non-spherically symmetric Weyl-distorted spacetimes which nevertheless contain extremal spherically symmetric Reissner-Nordstr\"{o}m-type horizons can be found in \cite{Booth:2012rm}.}. 

There are also several interesting results which provide bounds using the Komar angular momentum $J\subs{K}$ 
(total angular momentum minus the matter contributions). Again assuming an axisymmetric and stable MOTS it has been demonstrated  
\cite{Dain:2011pi,Jaramillo:2011pg,Dain:2011kb} that for \emph{any} matter fields satisfying the dominant energy condition, the possible values 
of the Komar angular momentum are bound by the areal radius:
\be
| J\subs{K} | \leq \frac{R^2}{2} \; . \label{KomarBound}
\ee
This is saturated only for horizons whose geometries are isomorphic to extremal Kerr.  So this is another bound that should apply to MKN horizons. 
Quite recently this result has been extended to demonstrate that if the bound is nearly saturated then the horizon is necessarily nearly extremal Kerr
and even far from extremality the Kerr-family provides constraints on possible horizon geometries \cite{Reiris:2013jaa}. 

Using the MKN solutions as a concrete example, this paper will explore these bounds. 
%
%
% both the complications and structure that remains
%for non-asymptotically flat black holes. We will focus on how the horizons in these spacetimes satisfy the constraint (\ref{B2}) and
%extremality uniqueness theorems as well as how their geometries are bound by their proximity to extremality. 
% However, 
%we will also touch further on the strengths and weaknesses of proposed methods for defining energy 
%and regularizing surface gravity, Coulomb potential and angular velocity. 
%
The plan is as follows. In Section \ref{MKNSec} we introduce the MKN spacetimes and calculate
basic physical properties including horizon location, (non-unique) surface gravity, electric and magnetic charge and angular momentum. This is mostly a  review of existing knowledge.  Section \ref{RangeMKN} is the beginning of new
material. We show that for a fixed value of the magnetic field parameter one can uniquely identify MKN horizons by its areal radius, charge and angular momentum. 
We also demonstrate that the Harrison transform preserves the degree of extremality of the KN seed and finally consider the Komar angular momentum bound. 
Section \ref{Sec:HorizonGeom} explores the range of possible MKN horizon geometries and tests them against the bounds and constraints discussed above. 
%Section \ref{Sec:KomarMass} considers a Komar mass formula for these horizons and compares it to other proposed mass functions. 
Section \ref{Sec:Conclusion} concludes  the paper with a brief discussion. 
In  \ref{AppMKNfunctions} we give the explicit forms of several important (but lengthy) functions which  appear in the MKN metrics.

\section{Melvin-Kerr-Newman Spacetimes}
\label{MKNSec}

\subsection{Form of the solutions} 
\label{FormSol}
The Melvin-Kerr-Newman (MKN) family of solutions \cite{ernst:black,ernst:kerr} is described by a metric and electromagnetic 
potential respectively of the form
\begin{eqnarray}
ds^2  = \frac{1}{f} \left(-\Delta \sin^2  \! \theta  \mathrm{d} t^2  + H \sin^2  \! \theta \left( \frac{  \mathrm{d}r^2}{\Delta} +  \mathrm{d} \theta^2 \right) \right)  + f \left(|\Lambda_o|^2  \mathrm{d}\phi - \omega \mathrm{d}t \right)^2
\label{MKNbase} \, ,
\end{eqnarray}
and
\begin{equation}
A = A_t  \mathrm{d}t + | \Lambda_o |^2  A_\phi  \mathrm{d} \phi \, .  \label{ErnstA}
\end{equation}
Of the functions appearing in this solution, two are the same as for Kerr-Newman:
\bea
\Delta & = & r^2 -2mr + a^2 +q^2 \; \;  \mbox{and} \\
H & = & (r^2 + a^2)^2 - \Delta a^2 \sin^2  \! \theta \, ,
\eea
while  $f$, $\omega$ and $A_t$ and $A_\phi$ are different from KN and algebraically formidable. They are defined by 
the complex MKN Ernst potentials $(\mathcal{E}\subs{MKN}, \Phi\subs{MKN})$. For a general review of Ernst solutions see \cite{Griffiths:2009dfa}.
Here we will just define the parts necessary for studying the metric.  

First the potentials. These are  generated by a Harrison transform of the regular KN seed potentials 
$(\mathcal{E}\subs{KN}, \Phi\subs{KN})$. For KN
\bea
f\subs{KN} &=& \frac{H \sin^2  \! \theta}{\Sigma} \; , % \; \; \; \mbox{where} \; \; \;   A = (r^2 + a^2)^2 - \Delta a^2 \sin^2  \! \theta 
\eea
with $\Sigma = r^2 + a^2 \cos^2 \! \theta$ and the electromagnetic potential
\bea
 \Phi\subs{KN} & = & q \left( \frac{a- i r \cos \! \theta}{r + i a \cos \! \theta} \right) \; .  \label{Phi_KN}
 \eea
Along with the twist potential
\bea
\varphi\subs{KN} & = & \frac{2 a \cos \! \theta }{\Sigma} \left(- 2m (r^2 +a^2)  + (a^2 m-m r^2+q^2r) \sin^2  \! \theta \right) 
\eea
these define: 
\begin{equation}
\mathcal{E}\subs{KN} \equiv f\subs{KN} + \Phi\subs{KN} \bar{\Phi}\subs{KN} + i \varphi\subs{KN} \, . \label{mathcalE}
\end{equation}
In terms of the seeds, the MKN potentials are then
\bea
\mathcal{E} &=& \frac{\mathcal{E}\subs{KN}}{\Lambda}  \; \; \mbox{and} \label{E_MKN} \\
\Phi &=& \frac{\Phi\subs{KN} + \frac{1}{2} B \mathcal{E}\subs{KN}}{\Lambda}  \label{Phi_MKN}
\eea
for
\be
\Lambda =  1 + B \Phi\subs{KN} + \frac{1}{4} B^2 \mathcal{E}\subs{KN}  \, , \label{Lambda}
\ee
where $B$ is a free parameter. %and $\mathcal{E}$ and $\Phi$ retain their old Kerr-Newman values.

The defining functions for the MKN solutions are then obtained  from the potentials in the following way. 
First, as in (\ref{mathcalE}), $f$ is defined by the real part of $\mathcal{E}\subs{MKN}$:
\be
f = \mbox{Re} (\mathcal{E}) - \Phi \bar{\Phi}\, . 
\ee
The imaginary part is again the twist potential 
\be
\varphi = \mbox{Im}(\mathcal{E}) 
\ee
and defines $\omega$ via the system of differential equations
\bea
\partial_r  \omega & = &   - \frac{\sin \! \theta}{f^2} \left(\partial_\theta \varphi + i (\bar{\Phi} \partial_\theta \Phi - \Phi \partial_\theta \bar{\Phi}  )  \right) \; \mbox{and} \label{omegaEq} \\
\partial_\theta \omega & = & \frac{\Delta \sin \! \theta}{f^2} \left( \partial_r \varphi + i (\bar{\Phi} \partial_r \Phi - \Phi \partial_r \bar{\Phi} )\right) \; . \nonumber
\eea
The potential $\Phi$ is composed of the $\phi$-components of $A_a$ and 
$\tilde{A}_a$ (the ``dual'' vector potential that generates $\star F$):
\be
\Phi \equiv A_\phi + i \tilde{A}_\phi \label{tA} \, . 
\ee
Thus the $t$ component of  (\ref{ErnstA}) is also defined by differential equations:
\begin{eqnarray}
\partial_r A_t & = & - \frac{1}{2f} \left(\Delta \sin \! \theta \partial_r \tilde{A}_\phi + 2 f \omega \partial_\theta {A}_\phi  \right)  \label{Adef} \; \; \mbox{and} \label{Aeq} \\
\partial_\theta A_t & = & \phantom{-} \frac{1}{2f} \left(\sin \! \theta \partial_\theta \tilde{A}_\phi - 2 f \omega \partial_r A_\phi  \right) \, . \nonumber
\end{eqnarray}
Finally, 
\be
|\Lambda_o|^2 =    |\Lambda_{\theta =0}|^2 = 1+ \left(\frac{3q^2}{2} \right)B^2 + \left( 2amq\right) B^3+  \left(a^2m^2 +  \frac{q^4}{16} \right)B^4\, . \label{Lambda021}
\ee
This factor is included to eliminate conical singularities from the metric. 

Equations (\ref{omegaEq}) and (\ref{Aeq}) can be solved and 
$f$, $\omega$ and $A_a$ written out explicitly but they are complicated enough (see for example \cite{Gibbons:2013yq}) 
that the forms are not very helpful and most direct calculations with the explicit metric are not practical. 
Luckily, it turns out that a knowledge of the Ernst potentials is sufficient for many calculations, including most of the quantities
used in this paper. For a few calculations, values of some of metric components and electromagnetic potential at the horizon are required  and these are
given in  \ref{AppMKNfunctions}.

If we set $m=a=q=0$, this spacetime is the Melvin magnetic universe\cite{Bonnor:1953xx,Bonnor:1954xx,Melvin:1964xx}. The MKN solutions can be thought of as black holes immersed in background
 magnetic fields, where $B$ parameterizes the strength of the magnetic field. That said the physical interpretation is not completely straightforward as these
 spacetimes are neither asymptotically flat nor even asymptotically Melvin\cite{Hiscock:1980zf,Gibbons:2013yq}. 
 Since we focus on quasilocal quantities and horizons this is not a major concern for us. 
% However, it should be kept in mind that the classical definition of a black hole
% is for asymptotically flat spacetimes as are definitions of global physical quantities such as the ADM mass.
% We will return to this issue in Section \ref{Sec:KomarMass}.  
 
 \subsection{MKN Horizons}
\label{Sec:GeometricProperties}

In this subsection we locate MKN horizons and study their basic geometric properties. 

\subsubsection{Killing horizon and surface gravity} 
\label{KHSG}

Given the asymptotics of the MKN spacetimes, they do not contain event horizons. However, if there is an event horizon at $r=r\subs{H}$ in a seed KN solution, 
it remains as a Killing (and hence marginally outer trapped) horizon for all MKN solutions generated from it via the Harrison transform.
That is $r=r\subs{H}$ for all values of $B$. The associated Killing vector field is 
\be
\xi_o = \frac{\partial}{\partial t} + \frac{\Omega\subs{H}}{| \Lambda_o|^2} \frac{\partial}{\partial \phi}  \; , \label{KV}
\ee
where $\Omega_H$ is the constant value of $\omega$ as evaluated on the horizon (see \ref{AppMKNfunctions} for the explicit form). 
%
%The norm of this vector field is given by
%\be
%\| \xi_o \|^2 = \frac{1}{f\subs{MKN}}  \left(   f^2_{\mbox{\tiny{MKN}}} \left( \omega_{\mbox{\tiny{MKN}}} - \Omega\subs{H} \right)^2 
%-  \Delta \sin^2 \! \theta \right) \, , \label{KillVan}
%\ee
%and clearly this vanishes for $r =r\subs{H}$ and will generally not vanish for $r \neq r\subs{H}$\footnote{See 
%\cite{Gibbons:2013yq} for an extended discussion of where $\xi$ is timelike or spacelike and the intricacies of trying to 
%define ergoregions for this spacetime.}. Hence $r\subs{H}$ is a Killing horizon. 

We consider the geometry of this horizon. On two-dimensional cross-sections the induced metric is
\be
dS^2 = 
% \left( \frac{H \sin^2 \! \theta}{f_{\mbox{\tiny{MKN}}}} \right)  \mathrm{d}\theta^2 + | \Lambda_o |^4 f_{\mbox{\tiny{MKN}}}  \mathrm{d} \phi^2 \, . % 
G\subs{H}(\theta) d \theta^2 + \left( \frac{| \Lambda_o |^4 (r\subs{H}^2+a^2)^2}{G\subs{H}(\theta)} \right) \sin^2 \! \theta d \phi^2 
 \label{InducedMetric}
\ee
%The position of the $\sin^2 \! \theta$ may appear unusual however note that $f\subs{MKN}$ (evaluated on the horizon) is of the form 
%\be
%f\subs{MKN} = \frac{(r^2+a^2)^2 \sin^2 \! \theta}{G\subs{H}(\theta)}
%\ee
where $G\subs{H}(\theta)$ is a quite complicated function (\ref{AppMKNfunctions} again).

This metric, of course, fully determines the intrinsic geometry including the area element which does not depend on $G\subs{H}(\theta)$:
\be
\tilde{\epsilon}_{\mbox{\tiny{H}}} = | \Lambda_o |^2  (r_{\mbox{\tiny{H}}}^2 + a^2) \sin\! \theta  \mathrm{d}\theta \wedge  \mathrm{d} \phi \, . 
\label{AreaEl}
\ee
Hence, the area of the horizon is
\be
\mathcal{A}_{\mbox{\tiny{MKN}}}  = 4 \pi | \Lambda_o |^2 (r_{\mbox{\tiny{H}}}^2 + a^2)   \; ,  \label{AMKN}
\ee
and so relative to the KN seed solution 
$\mathcal{A}_{\mbox{\tiny{MKN}}}  = | \Lambda_o |^2 \mathcal{A}_{\mbox{\tiny{KN}}}$. 

The surface gravity associated with $\xi_o$ is even more closely related to that of the seed solution. By the standard methods for Killing horizons \cite{Wald} 
this can be calculated from the relation:
\be
\kappa^2 = - \frac{1}{2} (\nabla_a \xi^o_{b})(\nabla^a \xi_o^b) \, ,  \label{kappa2}
\ee 
whence:
\be
\kappa\subs{MKN} = \frac{1}{2 (r\subs{H}^2 + a^2)} \left. \frac{d \Delta}{d r}  \right|_{r\subs{H}} = 
\left( \frac{r\subs{H}-m}{r\subs{H}^2 + a^2} \right) =  \kappa\subs{KN} \, , \label{kappaMKN1} 
\ee
where $\kappa\subs{KN} $ is the surface gravity of the seed solution. For this scaling of the Killing vector field the surface gravity is unchanged, 
however it is important to keep in mind that  $\xi_o$ may be rescaled by any constant $\xi_o \rightarrow \alpha_o \xi_o$ and still be a suitable 
horizon-defining Killing vector field. %As will be discussed in more detail in Section \ref{Sec:KomarMass}
There is no natural way to fix this scaling;  unlike the KN family of spacetimes, 
the full MKN family is not asymptotically flat and so an
appropriate scaling cannot be read off from infinity. The surface gravity is only defined up to this freedom.

One characterization of an extremal horizon is one whose surface gravity vanishes. For this purpose the rescaling freedom does not matter: if we rescale zero 
by a constant then it is still zero. Thus the transformation of an extremal horizon is also extremal. %We will return to consider extremal horizons further in future sections.  

\subsubsection{Horizon as a marginally outer trapped surface}
\label{HMOTS}

Next let us consider the null expansions of the horizon. 
Surfaces of constant $t$ and $r$ have outward and inward oriented null normals:
\bea
\ell & = &  \frac{\partial}{\partial t} + \frac{\Delta}{\sqrt{A}} \frac{\partial}{\partial r} + \frac{\omega}{|\Lambda_o|^2} \frac{\partial}{\partial \phi} 
\label{ellN} \\
N & = &  \frac{f}{2\sin^2 \! \theta} \left(  \frac{1}{\Delta} \frac{\partial}{\partial t} - \frac{1}{\sqrt{A}} \frac{\partial}{\partial r} + \frac{\omega}{|\Lambda_o|^2\Delta} \frac{\partial}{\partial \phi} \right) \; .   \nonumber
\eea
Here the scaling has been carefully chosen so that the vectors remain geometrically well-defined at the horizon (even though 
the coordinate system itself fails there). In particular note that on the Killing horizon, the Killing vector field $\xi_o = \ell$. 

Then the associated outward expansion is
\be
\theta_{(\ell)} = \tilde{q}^{a b} \nabla_a \ell_b = \frac{\Delta H_r}{2 A^{3/2}}
\ee
for the inverse two-metric $\tilde{q}^{ab} = g^{a b} + \ell^a N^b + \ell^b N^a$. It is clear that this vanishes on the Killing horizon: as 
always the Killing horizon is a MOTS.

The inward expansion is
\be
\theta_{(N)} = \tilde{q}^{a b} \nabla_a N_b =  - \frac{A_r f}{4 H^{3/2} \sin^2 \! \theta} 
\ee
and on the horizon this takes the form
\be
\left. \theta_{(N)} \right|\subs{H} = - \left( \frac{4 r^4 +  a^2 (3r^2 - a^2 \sin^2 \! \theta - q^2 \sin^2 \! \theta - r^2 \cos^2 \! \theta) }{r(r^2+a^2) \,  G\subs{H}(\theta)} \right)
\ee
This is clearly negative everywhere for Melvin-Schwarzschild (MS) for which
\be
\left. \theta_{(N)} \right|\subs{H}^{\mbox{\tiny{MS}}} = - \frac{16}{r \left(4+  r^2 B^2 \sin^2 \! \theta  \right)^2}   \, , 
\ee
however a little algebraic analysis shows that this property continues to hold for all values of $(r,a,q,B)$.  Thus the inward expansion
of the Killing horizon is always negative and by the discussion of \cite{Booth:2007wu} the Killing horizon is a future outer 
trapping horizon \cite{Hayward:1993wb} with fully trapped surfaces ``just inside''. Equivalently in the language of 
\cite{Clement:2011np,Clement:2012vb} 
%Andersson:2005gq,Andersson:2007gy} 
the negative inward expansion implies that the horizon is stable. 
This is true for arbitrarily large $B$ and is what one would 
intuitively expect for a black hole horizon\footnote{Intuitively appealing as it is, this property does not always hold. There are 
Weyl-Schwarzschild spacetimes for which $\left. \theta_{(N)} \right|\subs{H}$ changes sign on 
the horizon\cite{Pilkington:2011aj}. Also, in highly dynamical spacetimes where horizons ``jump'' 
there can be MOTSs for which the fully trapped region lies outside rather than 
inside the surface\cite{Booth:2005ng}. In neither of these situations is the horizon stable. 
}. It will also be relevant in Sections \ref{NExt} and \ref{IntGeom} where we consider theorems which 
restrict the horizon geometry: these theorems require stable MOTS.  

\subsection{Physical properties}
\label{Sec:PhysProp}

We now turn to physical properties of the MKN horizons. The electric and magnetic charges are uniquely defined and there
are two common measures of angular momentum: the total angular momentum and the Komar angular momentum. In the 
next sections we will use them, along with surface area $\mathcal{A}$, to characterize our horizons. 
These are each calculated on spacelike two-dimensional surfaces $S$ that respect the symmetries of the spacetime.
%We treat them in this subsection. 

% along with a brief discussion of mass. 

%In these calculations, the outward and 
%inward (future) oriented null normals to $S$ will again be $\ell^a$ and $N^a$. They will be cross-scaled  so that $\ell \cdot N = -1$: 
%the $\ell$ and $N$ given in (\ref{ellN}) are examples of such normals. 
%The induced metric on $S$ is $\tilde{q}_{ab}$ and area form is $\tilde{\epsilon}$. Relative to the full four-volume form
%$\tilde{\epsilon}_{ab} =  \epsilon_{abcd}  \ell^c N^d$ or equivalently $\epsilon_{\underset{\leftarrow}{ab}}^{\phantom{ab}cd} = 2 \tilde{\epsilon}_{ab} N^{[c} \ell^{d]}$
%where the underarrow indicates a pull-back to $T_2^0 S$. When necessary we assume that both $S$ and the null normals 
%share the rotational symmetry of the spacetime.

\subsubsection{Electric and magnetic charges}
First, for any two-surface $S$, the contained electric and magnetic charges are:
\begin{eqnarray}
Q & \equiv & \frac{1}{4 \pi}  \int_S \! \! \tilde{F} = \frac{1}{4 \pi} \int_S  \! \!   \mathrm{d} \tilde{A} =   \frac{1}{4 \pi} \int_S  \! \!  \tilde{\epsilon} E_\perp \label{charges}  \\
P & \equiv &  \frac{1}{4 \pi}  \int_S \! \! {F} =  \frac{1}{4 \pi} \int_S  \! \!   \mathrm{d} {A} =   \frac{1}{4 \pi} \int_S  \! \!  \tilde{\epsilon} B_\perp \nonumber
\end{eqnarray}
where the two-forms are understood to be pulled back into $T_2^0 S$, $\tilde{\epsilon}$ is the induced area element on the 
two-surface (\ref{AreaEl}) and  $E_\perp = \ell^a N^b F_{ab}$ and $B_\perp =  \ell^a N^b \tilde{F}_{ab}$ 
are the normal components of the electric and magnetic fields. Note that these charges are determined directly by the components of 
the electromagnetic potential $\Phi$. 

Evaluating these for MKN the magnetic charge is seen to vanish ($P=0$) while \cite{AlievGaltsov}: 
\be
Q_{\mbox{\tiny{MKN}}}  = \frac{\left|\Lambda_o \right|^2}{2} \left[\tilde{A}_\phi^{\mbox{\tiny{MKN}}}  \right]_{\theta = \pi}^{\theta = 0} 
 =   q + 2 am B - \frac{1}{4}  q^3 B^2  \, .  \label{QMKN1}
\ee

\subsubsection{Angular momentum}
%MKN spacetimes are rotationally symmetric and so have well-defined angular momenta. There are two commonly 
%There are effectively two ways to do this and we will make use of both. 

%
%
%\footnote{There are two versions of the Komar angular 
%momentum: one is purely geometric and equivalent to integrating $\phi^a \tilde{\jmath}_a$ while the other is equivalent to integrating $\phi^a (\tilde{\jmath}_a + \tilde{\jmath}_a^{\tiny{EM}})$. For asymptotically flat spacetimes time with fall-off conditions on the matter they are equivalent however in our case they are not. Here we consider
%the second. The first is both harder to calculate and not so useful for our purposes.}, 
%

The total angular momentum includes matter contributions and has been derived by many different methods including those of Brown-York\cite{Brown:1992br} and 
isolated and dynamical horizons\cite{Ashtekar:2001is,Ashtekar:2001jb}. Calculated on a spacelike two-surface $S$ that respects the rotational symmetry:
\be
J \equiv \int_S \tilde{\epsilon} \phi^a (\tilde{\jmath}_a + \tilde{\jmath}_a^{\tiny{EM}}) \label{JAngMom}
\ee
where $\phi^a$ is the rotational Killing vector (scaled so that its integral curves have affine length $2 \pi$) and 
the geometric and matter contributions to the angular momentum are respectively:
\be
\tilde{\jmath}_a =  - \frac{1}{8 \pi} \tilde{q}_a^b {N}_c \nabla_b {\ell}^c   
\; \; \mbox{and} 
\; \; \tilde{\jmath}_a^{\tiny{EM}} = \frac{1}{4 \pi} E_\perp \tilde{A}_b
\ee
where $\ell^a$ and $N^a$ are null normals to $S$ as in Section \ref{HMOTS} and $\tilde{A}_b =  \tilde{q}_b^{\phantom{b} c} A_c$\, . 

Attempting to calculate this directly for the MKN spacetime in Boyer-Lindquist coordinates is not practical: the resulting expressions are too complicated
to handle even with the help of computer algebra. However, it can be calculated relatively easily from the Ernst potentials 
(or see \cite{Gibbons:2013dna} for an equivalent method):
\bea
J\subs{MKN}  &=&    \frac{| \Lambda_o |^4}{8} \left[\varphi + 2 A_\phi \tilde{A}_\phi \right]^{\theta=0}_{\theta=\pi} \label{JMKN1} \\
& = & am - q^3 B - \frac{3}{2}    q^2 am  B^2 -   q \left(2 a^2 m^2 + \frac{1}{4} q^4 \right) B^3  \nonumber  - am \left(a^2 m^2 + \frac{3}{16} q^4 \right)  B^4 \, . \nonumber
\eea
Then, as for the electromagnetic charges, this quantity is almost independent of the particular $S$ used for the calculation: if $S$ encloses the horizon 
$J=J\subs{MKN}$ but if it doesn't then $J=0$. 

In future sections, an unqualified ``angular momentum'' will always mean total angular momentum however occasionally we instead specify the Komar 
angular momentum\cite{Wald}. This is generally thought of as the angular momentum of the
gravitational fields alone and in terms of the quantities discussed above\footnote{The reader may be more familiar with an expression of the form
\begin{equation*}
J\subs{K} =\frac{1}{16} \int_{S} \! \tilde{\epsilon}  \left( \ell^a N^b \nabla_a \phi_b  \right) = \frac{1}{16} \int_{S} \! \! \star   d \phi \, ,
\end{equation*}
where here $d\phi$ is the exterior derivative of $\phi_a$.}: 
\be
J\subs{K} \equiv \int_S \tilde{\epsilon} \phi^a \tilde{\jmath}_a \, .  \label{JKomar}
\ee
Unlike the total angular momentum, the value of $J\subs{K}$ does depend on where it is evaluated. For the familiar case of asymptotically flat electrovac spacetimes
(like KN):
\be
J =  J\subs{K} [S_\infty] \, , 
\ee
where $S_\infty$ is a sphere at spacelike infinity: the electromagnetic terms fall off quickly enough that their contribution vanishes. 
However for more general asymptotics this is not necessarily the case. 
In particular for MKN the electromagnetic field does not drop off but instead grows in strength as one approaches infinity. 

The Komar angular momentum for MKN does not evaluate to a simple form like (\ref{JMKN1}). We evaluate it on the horizon (the value needed for the theorems 
that we will consider in later sections)  however even there things are not simple. On a cross-section of the Killing horizon $S\subs{H}$:
\be
J\subs{K} [S\subs{H}] = \frac{|\Lambda_0|^2 (r\subs{H}^2+a^2)^4}{8}  \int_0^\pi \left( \frac{\sin^3 \theta \partial_r \omega}{G\subs{H}^2(\theta)} \right) \rmd \theta
\ee
where $\partial_r \omega$ may be calculated from (\ref{omegaEq}). While this integral 
can be evaluated in closed form, the result is complicated enough that it is not useful to present it explicitly (even after significant simplification the expression
is well over a page long). In future sections we will evaluate this numerically when necessary.

\section{Range of MKN spacetimes}
\label{RangeMKN}

In this section we consider the range of the physical parameters for the spacetimes generated by the Harrison transformation. 

In doing this it is useful to work with the measurable physical parameters $(R, Q, J)$ which, unlike metric parameters $(m,q,a)$, are 
well defined for both KN and MKN spacetimes. For
KN solutions:
\be
R = \sqrt{r\subs{H}^2 + a^2} \; , \; \; Q=q \; \mbox{and} \; \; J=am \, ,  \label{KNPar}
\ee
 where $r\subs{H}$ is the positive root of $\Delta(r) = r^2 -2 m r + a^2 + q^2 = 0$. Conversely
\be
m = \frac{\mathscr{R}^2}{2R} \; , \;  \; r\subs{H} = \frac{(R^2+Q^2)R}{\mathscr{R}^2}  \; , \; \; 
a = \frac{2JR}{\mathscr{R}^2} \label{a0}  \; \; \mbox{and} \; \; q  =  Q \, , \label{ConvTrans}
\ee
for $\mathscr{R}^2 = \sqrt{(R^2+Q^2)^2 + 4 J^2} $.

The effects of the Harrison transformations can then be rewritten as: 
\bea
R^2_{\mbox{\tiny{MKN}}}  &=& \left( 1+ \left(\frac{3Q^2}{2} \right)B^2 + \left( 2JQ\right) B^3+  \left(J^2 +  \frac{Q^4}{16} \right)B^4 \label{R__MKN} \right) R^2\, ,  \label{R_MKN} \\
Q_{\mbox{\tiny{MKN}}} &=& Q + 2 JB - \frac{1}{4}  Q^3 B^2 \label{Q_MKN} \; \; \mbox{and} \\
J_{\mbox{\tiny{MKN}}} &=& J - Q^3 B - \frac{3}{2} J Q^2 B^2 - Q \left(2 J^2 + \frac{1}{4} Q^4 \right) B^3 \label{J_MKN}\\
& & - J \left(J^2 + \frac{3}{16} Q^4 \right)  B^4 \, .  \nonumber 
\eea
To emphasize that these equations give the mapping between the measurable characteristics of the seed and final solutions, we refer to them as the 
\emph{physical parameter transforms}. Then we can demonstrate the following important properties. 

\subsection{Physical parameters are sufficient to specify a unique MKN horizon}
\label{1-1}
In this section we show that a set of values $(B_o, R_o, Q_o, J_o)$  is sufficient to specify a unique MKN horizon, just as $(R_o, Q_o, J_o)$ is sufficient for KN. 
%
%In this subsection we demonstrate that for any fixed value of $B$ the physical parameter transforms are bijective. Thus just as a set of values
%$(R_o, Q_o, J_o)$ is sufficient to specify a unique KN horizon, a set $(B_o, R_o, Q_o, J_o)$ specifies a unique MKN horizon.

% Further 
%
%
%there is a natural bijective mapping between KN and MKN horizons. This mapping is defined by 
%matching horizons with the same physical parameters and provides a natural comparison when studying MKN geometries.
%%but also confirms that the bound (\ref{B2}) holds for MKN horizons. 

As defined in Section \ref{FormSol}, MKN spacetimes (and so horizons) are a four-parameter family of spacetimes
specified by $(B,m,q,a)$. They can equivalently be specified by $(B,R\subs{seed},Q\subs{seed},J\subs{seed})$ but those are parameters 
for the seed rather than MKN horizon. Thus to see that $(B_o, R_o, Q_o, J_o)$ is sufficient we need to demonstrate that the set of equations
\be
R\subs{MKN} = R_o  \;  , \; \;  Q\subs{MKN} = Q_o \; \mbox{and} \; \;  J\subs{MKN} = J_o \label{Peq}
\ee
has a unique solution $(R,Q,J)$ for each $B_o$. 

We begin by showing that a solution always exists. First, it is obvious that (\ref{R_MKN}) fixes $R$ given a $Q$ and $J$ and so the 
proof rests on (\ref{Q_MKN}) and (\ref{J_MKN}). The result is trivial for $B=0$ and so we restrict our attention to $B \neq 0$. Then we can use $B$ to fix a length
scale and so remove it from the equations. For $Q=\tilde{Q}/B$ and $J=\tilde{J}/B^2$, (\ref{Q_MKN}) can be solved for $\tilde{J}$ as:
\be
\tilde{J} = \frac{1}{2} \left(\frac{1}{4} \tilde{Q}^3 - \tilde{Q} + \tilde{Q}_o  \right) \, . \label{tJ}
\ee
Substituting this in (\ref{J_MKN}) we get a ninth-degree polynomial equation in $\tilde{Q}$:
\be
F(\tilde{Q}) = 0 
\ee
where
\bea
F(\tilde{Q}) & = &  \left( \frac{1}{512} \right) \tilde{Q}^9 + \left( \frac{1}{32} \right) \tilde{Q}^7 +\left( \frac{3 \tilde{Q}_o}{128} \right) \tilde{Q}^6 + \left( \frac{3}{16} \right) \tilde{Q}^5 
 + \left(\frac{5 \tilde{Q}_o}{32}\right) \tilde{Q}^4 \nonumber \\
 & & +  \left( \frac{3\tilde{Q}_o^2}{32} + \frac{1}{2} \right) \tilde{Q}^3 + \left( \frac{\tilde{Q}_o }{8} \right) \tilde{Q}^2 
 + \left( \frac{\tilde{Q}_o^2 }{8} + \frac{1}{2} \right) \tilde{Q}   \label{FQ}  \\ 
 & &   + \left(\frac{ \tilde{Q}_o^3}{8} - \frac{\tilde{Q}_o}{2} +  J_o  \right) \, .  \nonumber
\eea
Given that $F$ is of odd order, $F(\tilde{Q}) = 0$ has at least one solution and so the equations (\ref{Peq}) always have a solution.
 
 It remains to show that it is unique. To see this note that
\be
\frac{dF}{d\tilde{Q}} = \frac{1}{512} (9 \tilde{Q}^4 + 4) \left(16 \tilde{Q}_o^2 + 8 \tilde{Q} [\tilde{Q}^2 + 4 ] \tilde{Q}_o + [\tilde{Q}^2 + 4 ]^3  \right) 
\ee
where we have regrouped the term in the right-most parentheses as a quadratic polynomial in $\tilde{Q}_o$. The discriminant of that term is
everywhere negative and so $\frac{dF}{d\tilde{Q}}$ is nowhere vanishing. Hence since it is positive when $\tilde{Q}=0$, it is positive everywhere. Thus 
$F(\tilde{Q})$ is monotonically increasing and the mappings (\ref{R_MKN}-\ref{J_MKN}) are one-to-one from $(R,Q,J)$ to $(R\subs{MKN},Q\subs{MKN},J\subs{MKN})$ and we are done.

\subsection{Harrison transform preserves the degree of extremality}

The results of subsection \ref{1-1} depend only on the form of the transformations and apply for any real $(R_o,Q_o,J_o)$. However, the parameters are not
entirely independent. For KN horizons:
\be
a^2 + q^2 \leq m^2 \;  \Longleftrightarrow \; Q^4 + 4 J^2  \leq R^4 
\ee
where the inequality is saturated for extremal horizons.% (and violated for naked singularities). 

As noted in the introduction, it has recently been proved that this inequality is a \emph{universal bound} for stable MOTS in four-dimensional electrovac spacetimes\cite{Clement:2011np,Clement:2012vb} (though it can be violated in anti-deSitter spacetimes \cite{Booth:2007wu}  or in higher dimensions \cite{Myers:1986un,Emparan:2003sy}). Thus we can define an \emph{extremality parameter}
\be
\chi^2 =  \frac{4 J^2 + Q^4}{R^4}
\ee
such that $0 \leq \chi^2 \leq 1$ with any $\chi^2 = 1$ horizon said to be extremal. This bound should equally well apply to MKN horizons 
and in fact this is easy to see. 
By direct calculation
\be
4 J_{\mbox{\tiny{MKN}}}^2 + Q_{\mbox{\tiny{MKN}}}^4 = | \Lambda_o |^4 \left(4 J^2 + Q^4  \right) \, , 
\ee
and so 
\be
\chi^2\subs{MKN} = 4 \mathcal{J}\subs{MKN}^2 + \mathcal{Q}^4\subs{MKN}  = 4 \mathcal{J}\subs{seed}^2 + \mathcal{Q}^4\subs{seed} = \chi^2\subs{seed} \, ,
\label{ExInvariant}
\ee
where we have adopted the convention that a calligraphic quantity is the dimensionless version of the corresponding physical property as scaled against  
$R\subs{seed}$ or $R\subs{MKN}$ (as indicated by the subscript). Thus the Harrison transformation preserves the degree of extremality and in particular  this 
confirms that the KN bound (\ref{B2}) also holds for MKN horizons. 

%
%Sub-extremal KN seed solutions give rise to sub-extremal MKN solutions while 
%extremal KN solutions are transformed into extremal MKN solutions. 
%Note too that if $(Q,J)  = (0,0)$ then for all $B$, $(Q\subs{MKN}, J\subs{MKN})= (0,0)$: Schwarzschild maps into Melvin-Schwarzschild.  

\begin{figure}
\begin{center}
\includegraphics{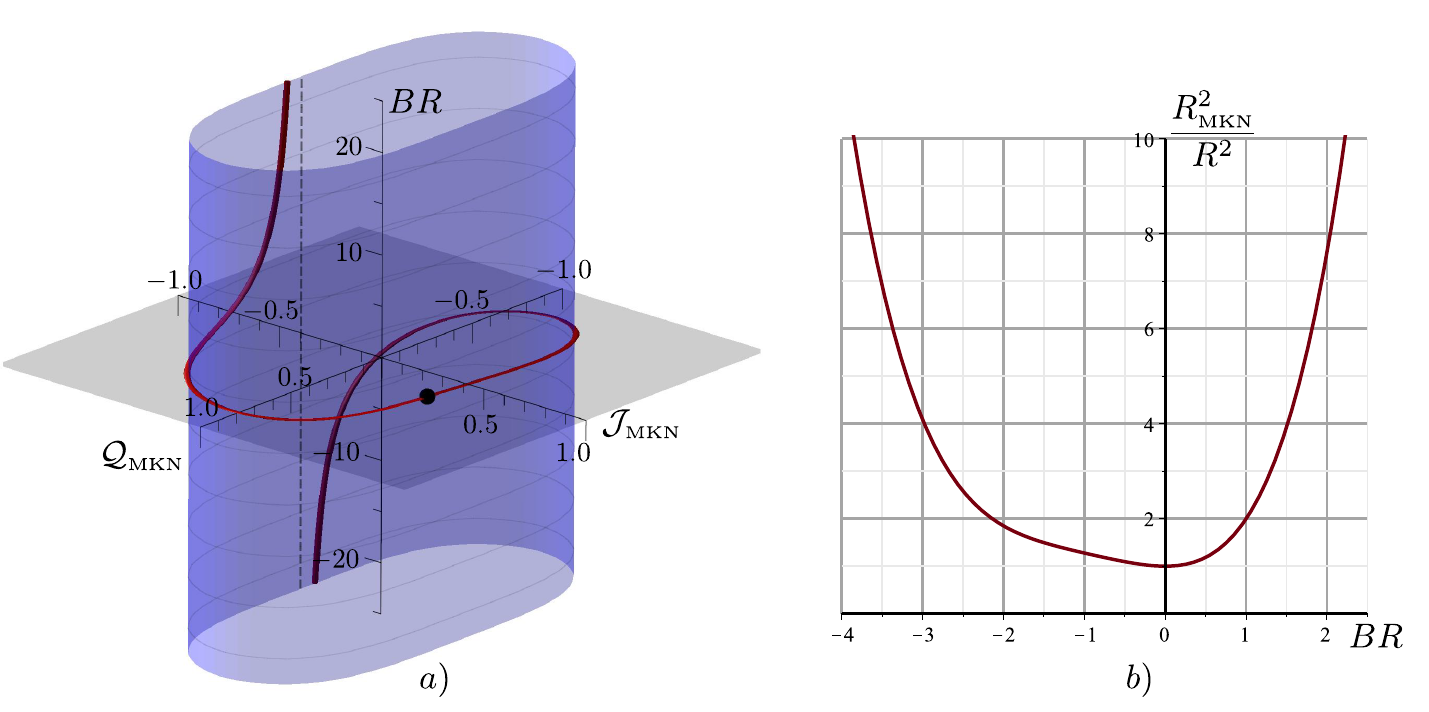}
\end{center}
\caption{Effect of Harrison transform on physical properties of MKN solutions. \\
 a) On the left the black dot marks the physical 
properties of the seed solution $Q= 0.2R$ and $J = 0.4R$ while the red line records how they evolve for changing $B$. 
The dark purple surface around which it wraps is a surface of constant $\chi^2$ and the dashed line in back shows the asymptotic value of the properties as $B \rightarrow \pm \infty$. 
\\b) On the right for the same values of $(Q,J)$, areal radius increases monotonically with the magnitude of $B$.  }
\label{HarTr}
\end{figure}

We can make use of this invariance to better understand the physical parameter transforms. Figure \ref{HarTr}a) shows how the
transformation changes the physical properties of the solution as $B$ runs from $-\infty$ to $\infty$. A particular seed solution is shown, however the behaviour is
generic: the evolution curve wraps around the $\mathcal{Q}^4+4 \mathcal{J}^2 = \chi^2$ surface and ultimately asymptotes to 
\begin{eqnarray}
{\mathcal{Q}}_\infty &=& \lim_{B \rightarrow \pm \infty}  {\mathcal{Q}}\subs{MKN}  = -\frac{\mathcal{Q}\subs{seed}^3}{\sqrt{16\mathcal{J}^2\subs{seed}+\mathcal{Q}\subs{seed}^4}} \label{Qa} \\
{\mathcal{J}}_\infty &=& \lim_{B \rightarrow \pm \infty}  {J}\subs{MKN} =  - \mathcal{J}\subs{seed}  \left( \frac{16\mathcal{J}^2\subs{seed}+3 \mathcal{Q}\subs{seed}^4 }{16\mathcal{J}^2\subs{seed}+\mathcal{Q}\subs{seed}^4 } \right) \label{Ja}
\end{eqnarray}
from both directions. 

At the same time, from (\ref{R_MKN}) it follows that 
$R\subs{MKN}$ has an absolute minimum at $B=0$:
\be
\frac{d (R^2\subs{MKN})}{dB} = \frac{B}{4} \left(Q^2 (3 + BQ^2) + (4BJ+3Q)^2 \right) \, . 
\ee
If either of $Q$ or $J$ are non-vanishing then this vanishes only for $B=0$, is negative for $B < 0$ and positive for $B>0$. Thus area increases monotonically 
with the magnitude of $B$ and goes to infinity as $B \rightarrow \pm \infty$. This is shown in Figure \ref{HarTr}b) for the same seed solution as earlier and again 
the behaviour is qualitatively generic. The only exception is $Q=J=0$ (Melvin-Schwarzschild) for which the area is independent of $B$.   

\subsection{Komar angular momentum is bound by area}

Recall \cite{Dain:2011pi,Jaramillo:2011pg,Dain:2011kb} that there is also a universal bound on the Komar angular momentum (\ref{KomarBound}). To confirm this
for the MKN horizons we need to demonstrate that
\be
\frac{|J_{\mbox{\tiny K}}|}{R\subs{MKN}^2} \leq \frac{1}{2} \, . \label{KB2}
\ee
This is not completely trivial even for pure KN. For $B=0$:
\bea
\frac{J\subs{K}}{R^2} = &\frac{1}{R^2}& \left( J -  \frac{Q^2((Q^2+R^2)^2-4J^2)}{8J(Q^2+R^2)} \right. \\
& &\left.  - \frac{Q^2 ((Q^2+R^2)^2+4J^2)^2}{16J^2(Q^2+R^2)} \arctan \left(\frac{2J}{Q^2+R^2} \right) \right)\; . \nonumber
\eea
Scaling out $R$ and plotting this for $|\mathcal{Q}| \leq 1$ and $|\mathcal{J}| \leq \frac{1}{2}  \sqrt{R^4-Q^4}$ it can be seen that 
(\ref{KB2}) holds and is saturated only for $Q=0$ and $J=\frac{1}{2}$. This is in agreement with the theorems. 

For MKN horizons the situation is much more complicated and not easily presented: even after extensive algebraic simplification the 
expression for $J\subs{K}$ is well over a page long
and now depends on three variables $(\mathcal{Q}, \mathcal{J}, \mathcal{B})$ (with $R$ scaled out). Though we have not succeeded in analytically
confirming (\ref{KB2}) in this case, extensive numerical investigations were all consistent with the inequality. For very large $\mathcal{B}$ the inequality 
is always far from being saturated: asymptotically in  $\mathcal{B}$
\be
%\lim_{\mathcal{B} \rightarrow \infty} 
\frac{|J_{\mbox{\tiny K}}|}{R\subs{MKN}^2} \approx O \left(\frac{1}{\mathcal{B}^4} \right)\, ,
\ee
and so goes to zero as $\mathcal{B} \rightarrow \pm \infty$. In fact (\ref{KB2}) is only saturated for pure extremal 
Kerr: $(\mathcal{Q}=0,\mathcal{J}=\frac{1}{2},\mathcal{B} = 0)$. It is never saturated for $\mathcal{B} \neq 0$. 

\section{Horizon geometry}
\label{Sec:HorizonGeom}

In this section we examine the geometry of MKN horizons which, in comparison to KN horizons, range from highly distorted
for MS ($\chi = 0$) to isometric for extremal MKN ($\chi = 1$). We begin with the extremal case. 

%We begin with the extremal case $\chi = 1$ which is 
%strongly constrained by the extremal horizon uniqueness theorems. 

%We now consider how these observations extend to general MKN horizons. 

\subsection{Extremal MKN ($\chi = 1$)}
\label{extremality}

%While $\chi=0$ horizons can be arbitrarily distorted relative Schwarzschild solutions of the same area, things are very 
%different for $\chi = 1$. 
%
%
%Extremal horizons not only satisfy $Q\subs{MKN}^4 + 4 J\subs{MKN}^2 = R\subs{MKN}^4$ but also have vanishing surface gravity (\ref{kappaMKN1}). 
%As noted earlier, t
There are well-known theorems which prove that any four-dimensional electrovac extremal isolated horizon is necessarily isomorphic to an  extremal horizon in the Kerr-Newman family\cite{Hajicek:1974oua, Lewandowski:2002ua, Kunduri:2007vf}.
This isomorphism includes  an isometry of the cross-sections of the horizon as well as identical angular-momentum one-forms 
$(\tilde{\jmath }_a + \tilde{\jmath}^{\mbox{\tiny{EM}}})$, and pull-backs of 
$F_{ab}$ and $\tilde{F}_{ab}$ into the surface. Consequently  the areal radius, angular momentum and electric and magnetic charges 
will also match those of the corresponding extremal Kerr-Newman horizon. 
This result does not depend on global properties of the spacetime, including asymptotic structure, 
and so certainly applies to the MKN spacetimes. 

For MKN horizons, we can algebraically demonstrate this identity. Here we just consider the geometry however similar
calculations could be done for the other properties. 
%
% at the level of the geometry (the other properties could
%be similarly checked). 
From  (\ref{InducedMetric}) the induced metric on an extremal MKN cross-section is
\be
dS^2 = G\subs{ex}  \mathrm{d}\theta^2 +   | \Lambda_o |^4 \left(\frac{(r^2+a^2)^2 \sin^2 \!\theta}{G\subs{ex}} \right)  \mathrm{d} \phi^2 \, ,
\ee
where from \ref{AppMKNfunctions} (after selectively substituting $r^2 = m^2 = a^2 + q^2$ and doing some work to
find the simplest possible form for the final expressions) we find
\be
G\subs{ex} = m\subs{MKN}^2 + a\subs{MKN}^2 \cos^2 \! \theta \label{Gex}
\ee
where 
\bea
m\subs{MKN} & = & m + a q B + \frac{1}{4} m \left( 4 a^2 + q^2\right) B^2 \label{mMKN}\\
a\subs{MKN} & = & a - qm B - \frac{1}{4} a\left( 4 a^2 + 3q^2 \right) B^2 \label{aMKN}
\eea
Note that we have left some $m$s in these expressions to avoid writing square roots\footnote{This mapping was also observed independently in 
\cite{Hejda:2015gna,Bicak:2015lxa}.}. 

The choice of variable names and disappearance of $\cos^4 \! \theta$ terms in (\ref{Gex})
is of course no coincidence: by the uniqueness theorems the induced metric has to be of the same form as for
some extremal KN. Thus $m\subs{MKN}$ and $a\subs{MKN}$ are the metric parameters for the extremal KN horizon 
which matches up with the MKN horizon. Combining this with (\ref{Q_MKN}):
\bea
q_{\mbox{\tiny{MKN}}} &=& q + 2 am B - \frac{1}{4}  q^3 B^2  \label{qMKN}
\eea
we have a specialized set of mappings that explicitly demonstrate how the Harrison transform maps extremal KN horizons into extremal MKN horizons.
It is a pleasant surprise to find that they are simply quadratic in $B$ (as opposed to the quartic expressions for $R\subs{MKN}$ and  $J\subs{MKN}$). 
As a consistency check it is straightforward to show that they reproduce (\ref{Q_MKN}-\ref{R_MKN}) if those equations are restricted to the extremal case. 

Of course such a mapping of metric parameters \emph{cannot} be found in general. As will be explicitly demonstrated in the next few subsections, 
non-extremal MKN horizons (with $B \neq 0$) are not members of the standard KN family.

\subsection{Near extremal geometries: $\chi \lesssim 1$}
\label{NExt}

There are also ``near-uniqueness'' theorems for near-extremal horizons. In \cite{Reiris:2013jaa}, Reiris and 
Gabach-Clement proved a series of results constraining axisymmetric black hole horizons relative to that of
extremal Kerr. In particular they showed that for spacetimes satisfying the dominant energy condition there are universal bounds for how far the horizon
geometry and angular momentum one-form (expressed in their results as a potential) may deviate from extremal Kerr. 
They also quantitatively showed how those properties must approach extremal Kerr as the Komar angular momentum $J\subs{K} \rightarrow \frac{1}{2} R^2$. 

Their results hold for our electrovac MKN spacetimes however we note that with the inclusion
of charge they necessarily become looser: for example Reissner-Norstr\"om with $Q=R$ is certainly extremal but for this horizon
$J\subs{K} = 0$. Thus, while they track proximity to Kerr extremality with the 
parameter 
\be
\delta\subs{RGC} = 2 \sqrt{\frac{R^4}{4J\subs{K}^2}-1} \,  , \label{deltaRGC}
\ee
 for our charged and rotating horizons it is more natural to  track proximity to KN extremality. As such we define the analogous
 \be
 \delta = \frac{2}{\chi} \sqrt{1 - \chi^2} = 2 \sqrt{\frac{R^4}{Q^4+4J^2}-1 }  \, . 
 \ee 
This is equivalent to (\ref{deltaRGC}) for pure vacuum horizons.% though keep in mind that we are now using the total rather than Komar angular momentum. 
 
As in the previous section we restrict our attention to horizon geometry, leaving aside the angular momentum one-form and 
electromagnetic field components. In \cite{Reiris:2013jaa} comparisons are made between a horizon and the extremal 
horizon with the same angular momentum. With the inclusion of charge the equivalent comparison is the MKN horizon with
$(R\subs{MKN},Q\subs{MKN},J\subs{MKN})$ versus the extremal KN horizon characterized by
$( 
%\sqrt[\leftroot{-1}\uproot{2}4]{Q\subs{MKN}^4+4J\subs{MKN}^2}
\sqrt{\chi} R\subs{MKN},Q\subs{MKN},J\subs{MKN})$. 

Adopting the notation of \cite{Reiris:2013jaa} we write horizon metrics in areal form
\be
dS^2 = R\subs{MKN}^4 \left( \frac{G\subs{H}}{R\subs{MKN}^4} \right) \rmd \theta^2 + \left( \frac{R\subs{MKN}^4}{G\subs{H}} \right) \sin^2 \!  \theta \rmd \phi^2 \, ,
\ee
and compare the areal radii and $e^\sigma = \frac{R\subs{MKN}^4}{G\subs{H}}$. 

Rewriting both as functions of $(\delta, \mathcal{J}, \mathcal{B}, \theta)$ ($R$ may be scaled out) we can expand them around $\delta=0$. The
areal radius ratio is easy:
\be
\frac{R\subs{MKN}}{R\subs{extreme KN}} = \frac{1}{\sqrt{\chi}} = \sqrt[\leftroot{-1}\uproot{2}4]{1+\frac{\delta^2}{4}} =
1 + \frac{\delta^2}{16} + O(\delta^4) \, . 
 \ee
The second involves $G\subs{H}$ and is algebraically very complicated.  It is not very useful to 
present it here in closed form, however after (numerically) extremizing over $\theta$, $\mathcal{J}$ and $\mathcal{B}$ 
it is not hard to show that:
\be
 1 - \frac{\delta^2}{8} + O (\delta^4) \leq \frac{e^{\sigma\subs{MKN}}}{e^{\sigma\subs{extreme KN}}}  \leq 1+ \frac{\delta^2}{8}
 + O (\delta^4) 
\ee 
with the lower bound attained for $\theta= \frac{\pi}{2}$, $\mathcal{J}=0$, $\mathcal{B} \rightarrow \pm \infty$ and the 
upper for $\theta = 0, \pi$ (independent of $\mathcal{B}$ and $\mathcal{J}$).

Thus, even for arbitrarily large $\mathcal{B}$ the near-extremal MKN horizons are tightly constrained by the extremal KN geometries. 
Far from extremality this is not the case.

\subsection{Melvin Schwarzschild ($\chi = 0$)}

%\begin{figure}
%\centering
%\includegraphics[scale=0.25]{EmbedKerr}
%\caption{Three Kerr horizons embedded in $\mathbb{R}^3$. The rotation parameters are: 
%a) $a/m = 0.2$, b) $a/m=0.6$ and c) $a/m=0.863$ (for which the Ricci scalar at the poles is on the verge of becoming negative). 
% All axes are in units of $m$. }
%\label{fig:KBHs}
%\end{figure}
%

%It is well-known that turning on angular momentum causes an initially spherical Schwarzschild horizon to become a progressively more oblate KN horizon 
%(Figure \ref{fig:KBHs}) until for 
%\begin{equation}
%\alpha^2 \equiv \frac{a}{r\subs{H}} = \frac{4 J^2}{(R^2 + Q^2)^2} > \frac{1}{3} 
%\end{equation}
%the two-dimensional Ricci scalar becomes negative at the pole. From that point up to (and including) extremality ($\alpha^2 =1$), 
%the horizon is no longer embeddable in Euclidean $\mathbb{R}^3$. 

MKN spacetimes with $\chi = 0$ are MS  and it is straighforward to see that increasing $B$ causes 
these horizons to become more prolate. 
To see this note that for $\chi = 0$, the metric on the horizon (\ref{InducedMetric}) takes the 
particularly simple form:
\be
dS^2 = R^2 \left(\lambda^2  \mathrm{d} \theta^2 + \frac{\sin^2 \! \theta}{\lambda^2}  \mathrm{d} \phi^2 \right) 
\ee
where
\be
\lambda^2 = \frac{G\subs{H}}{R^2} = \left( 1+ \frac{ \mathcal{B}^2 \sin^2 \theta}{4} \right)^2 \, .
\ee
%and the calligraphic $\mathcal{B} = BR$ is scaled in the usual fashion. 
The horizon geometry can then be seen to have several characteristics.

\subsubsection{Prolateness:} From the metric it is clear that while the horizon area is independent of $\mathcal{B}$ the 
overall geometry is not. As $\mathcal{B} \rightarrow \pm \infty$ the length of the prime meridian similarly diverges 
while the length of any parallel of latitude goes to zero.
The horizon becomes infinitely long and infinitely thin as compared to the Schwarzschild horizon of the same area. 
Representative cases appear in Figure \ref{fig:MSHs}.

\subsubsection{Negative curvature at equator:}
\begin{figure}
\centering
\includegraphics{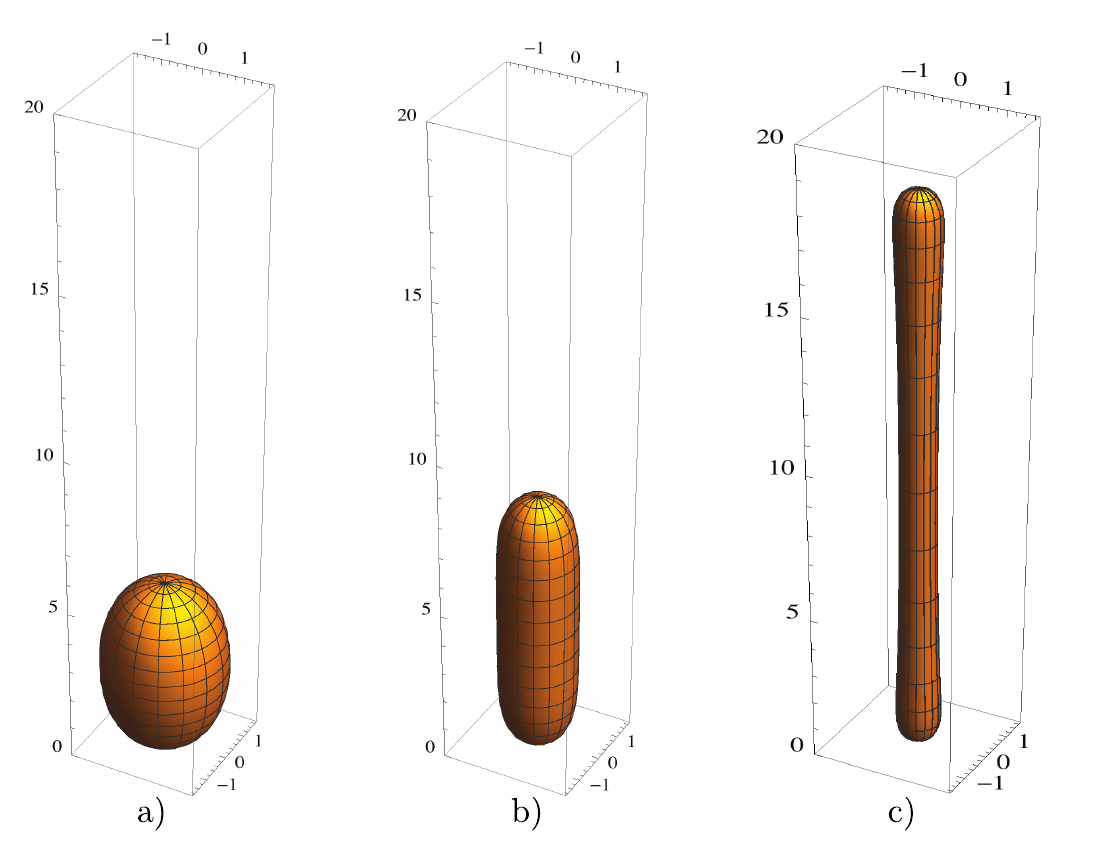}
\caption{Three MS horizons embedded in $\mathbb{R}^3$. The magnetic field parameters are:
a) $\mathcal{B} = 0.1$, b) $\mathcal{B}=2.0$ and c) $\mathcal{B}=4.0$. All axes are in units of $m=R/2$. }
\label{fig:MSHs}
\end{figure}

However, these horizons do more than just stretch out. For sufficiently large $\mathcal{B}$ they also develop an
hourglass shape, the occurrence of which is signalled by the development of a region of negative Gaussian 
curvature around the equator. 
This can be seen  in Figure \ref{fig:MSHs}, but can also be easily demonstrated from the two-dimensional Ricci scalar 
on the horizon:
\be
\tilde{R}\subs{MS} =  \frac{32\left( 16 + 32 \cos^2 \! \theta \mathcal{B}^2  + (3 \cos^4 \! \theta - 2 \cos^2 \! \theta -1  ) \mathcal{B}^4 \right) }{R^2 (4 + \mathcal{B}^2 \sin^2 \! \theta)^4} \, .
\ee
At the equator
\be
\tilde{R}\subs{MS} \left( \frac{\pi}{2} \right) = \frac{32(4-\mathcal{B}^2)}{R^2 (4 + \mathcal{B}^2)^3}
\ee
and so the horizon is hyperbolic there  for $|\mathcal{B}| > 2$. 

\subsubsection{Off-equator great circles:}
More details about the shape can obtained by considering the circumference of the parallels of latitude. We 
look for the longest parallels: great circles in the terminology of  \cite{Reiris:2013jaa}. The circumference 
function
\be
C(\theta) = \frac{8 \pi \sin \theta}{4 +  \mathcal{B}^2 \sin^2 \! \theta} \, ,
\ee
has extrema at $\theta=\frac{\pi}{2}$ and $\theta = \arcsin \left( \frac{2}{\mathcal{B}} \right)$. For $|\mathcal{B}|<2$ 
only the first is real and is a maximum: the horizon is convex in the usual way and the great circle is the equator. 
However for $|\mathcal{B}|>2$ the equator switches to becomes a minimum while the second pair of solutions are 
the maxima (the bulges of the hour-glass). As $\mathcal{B} \rightarrow \pm \infty$ the great circles move to 
the poles. 

\subsubsection{Great circle versus equator:}
For $|\mathcal{B}|>2$ the lengths of the equator and great circles are
\be
C\subs{eq} = \frac{8 \pi }{4 + \mathcal{B}^2} \; \; \mbox{and} \; \; C\subs{gc} = \frac{2 \pi}{|\mathcal{B}|} \, 
\ee
respectively. Thus while both are squeezed to zero length as $B \rightarrow \pm \infty$
\be
\lim_{B \rightarrow \pm \infty} \frac{C\subs{gc}}{C\subs{eq}} = \infty \, . 
\ee
The bulge circumference becomes arbitrarily large relative to the equator. 

Intuitively one might view this as the horizon becoming so stretched that the ends attempt to pinch off to form their 
own independent black holes in the manner of the Gregory-Laflamme instability\cite{Gregory:1993vy}.

\subsection{Intermediate geometries: $0 < \chi < 1$}
\label{IntGeom}

We now have a good understanding of the limiting cases. While for $\chi^2 = 0$ the horizon becomes infinitely stretched
and hour-glass shaped for $\mathcal{B} \rightarrow \pm \infty$, for $\chi^2 \approx 1$ it is always constrained to be very 
close to the corresponding extremal KN horizon. In this section we consider the intermediate region to see how it  
interpolates between these limits. In particular we will investigate the degree of distortion in the
$\mathcal{B} \rightarrow \pm \infty$ limits. These limits are also convenient to work with as they have 
relatively simple forms compared to the full complexity of finite $\mathcal{B}$. 

%
%Though exact expressions for the quantities considered in the previous section exist, they are 
%very complicated and so generally we will either only present the exact forms for limiting cases or 
%instead use graphs to demonstrate their behaviours. 
%
\subsubsection{Prolateness:}
We begin by demonstrating that in contrast to MS, general MKN solutions do not become 
infinitely stretched and squeezed in the $\mathcal{B} \rightarrow \pm \infty$ limit. 

From  (\ref{InducedMetric}) the prime meridian has length 
\be
L = \int_0^{\pi} \sqrt{G\subs{H}} d \theta   \label{LMKN}
\ee
In general this cannot be integrated in closed form  however
\be
\lim_{\mathcal{B} \rightarrow \pm \infty} \left( \frac{L}{\pi R\subs{MKN}} \right) = \frac{R^2}{\mathscr{R}^2 \sqrt{\mathcal{Q}^2+ 16 \mathcal{J}^2}}\int_o^\pi \mathcal{F} \rmd \theta 
\ee
where
\be
\! \! \! \! \! \! \! \!  \! \! \! \! \! \! \! \! \mathcal{F} = \sqrt{\bigg((1-\chi^2)\cos^2 \! \theta+(1+\mathcal{Q}^2+4\mathcal{J}^2) \bigg)^2+4\mathcal{J}^2 (\mathcal{Q}^2+2)^2 \cos^2 \! \theta} \, , \label{cF}
\ee
and we have scaled against $R\subs{MKN}$ since area increases for $\chi \neq 0$ (as seen in Figure \ref{HarTr}). 
If $\chi \rightarrow 0$ (so $\mathcal{Q}, \mathcal{J} \rightarrow 0$) then it is clear that this diverges as we saw earlier for MS. However
when $\chi \neq 0$ it is has a finite value: the horizon does not become infinitely stretched. 

At the same time the parallels of latitude have circumference:
\be
C(\theta) = \frac{2 \pi | \Lambda_o |^2 R^2 \sin(\theta)}{\sqrt{G\subs{H}(\theta)}} \, , \label{CMKN}
\ee
and
\be
\lim_{\mathcal{B} \rightarrow \pm \infty} \left( \frac{C(\theta)}{2 \pi R\subs{MKN}} \right) =  \left( \frac{\mathscr{R}}{R} \right)^2 \frac{\sqrt{\mathcal{Q}^4+16\mathcal{J}^2} \sin\! \theta}{\mathcal{F}} \, . 
\ee
Again if $\chi \rightarrow 0$ we recover the MS behaviour, however 
for $\chi \neq 0$ this limit does not vanish: the parallels are not squeezed to zero length.

\subsubsection{Negative curvature equator and poles:}
\begin{figure}
\centering
\includegraphics[scale=1]{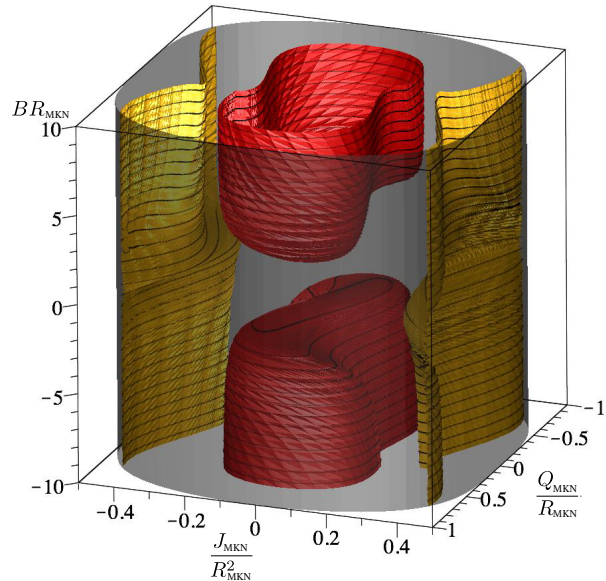}
\caption{Zeros of the Ricci scalar at the poles and equator of an MKN horizon. Each point in this plot corresponds to 
a class of MKN horizons (areal radius is scaled out). The charge and angular momentum on the
horizontal axes are the physical MKN parameters (as opposed to the unphysical parameters of the seed solutions). 
The set of MKN horizons whose Gaussian curvature vanishes at the equator is marked by the red surface while
the set whose Gaussian curvature vanishes at the poles is marked by the yellow surface. 
%
%The red surfaces marks the MKN horizons for which the Ricci scalar vanishes at the equator: inside that
%surface the curvature is negative and so the horizon has an hour-glass shape. The yellow surface marks the MKN %horizons for which the Ricci scalar vanishes at the poles: outside the surface the curvature is negative and so the horizon is 
%hyperbolic at the pole as for rapidly rotating KN . 
The translucent grey surface marks the set extremal horizons. }
\label{RicciScalar}
\end{figure}
Next we study the shape of the horizon using the Gaussian curvature. We demonstrate that there is only a finite portion 
of the $(\mathcal{Q}, \mathcal{J})$ phase space for which the horizon can develop an hourglass shape (even for 
arbitrarily large $\mathcal{B}$). Further in line with the expectation that nearly-extremal horizons are nearly KN, for 
sufficiently large $\mathcal{J}$ the horizon will always develop a region of negative curvature at the poles.

The two-dimensional Ricci scalar on the horizons is:
\be
\tilde{R} = \frac{2}{G\subs{H}} + \frac{1}{G\subs{H}^2} \left(  G''\subs{H} + 3 \cot \! \theta  \, G'\subs{H} \right) - \frac{2}{G\subs{H}^3} 
\left( G'\subs{H} \right)^2 \, ,
\ee
where primes indicate derivatives with respect to $\theta$. Writing out the (very complicated) general expression for this is 
not very useful so instead we present some of its properties in Figure \ref{RicciScalar}. That figure shows which regions
of the phase space of MKN horizons have negative curvature at either the equator or poles. 

First we consider the development of an hour-glass figure. In Figure \ref{RicciScalar}, the red surface marks the set of solutions whose horizons have vanishing Gaussian curvature at the equator. Inside
that surface all horizons have hyperbolic regions around the equator and so have the hour-glass figure seen for 
MS. By observation this can only happen for $\chi^2 < \frac{1}{4}$ and 
$\mathcal{B}>2$ (MS is the bound in this direction). 
Though this figure is restricted to $\mathcal{B} \leq 10$ the $\chi^2$ bound for the onset of hyperbolicity continues to hold 
as $\mathcal{B} \rightarrow \pm \infty$.
We discuss these bounds analytically in section 
\ref{OEGC_MKN}.

Next consider a characteristic of KN geometry: the development of negative Gaussian curvature at the
poles for sufficiently large angular momentum. In Figure \ref{RicciScalar}, the yellow surface marks the set of solutions
whose horizons have vanishing Gaussian curvature at the poles. Outside this surface all horizons have a region of 
negative curvature at the poles. Then it is clear that for sufficiently large $\mathcal{J}\subs{MKN}$ this always
develops. Again this continues to apply for arbitrarily large $\mathcal{B}$ and is in line with the expectations of 
\cite{Reiris:2013jaa} that close to extremality the geometry should be close to extremal KN. 

Note that there is no overlap of the regions of negative curvature: there is no MKN horizon which has negative 
curvature at both the poles and the equator. These are mutually exclusive geometric characteristics. Crudely, an MKN 
horizon can be either MS-like or KN-like but not both at the same time.

\subsubsection{Off-equator great circles}
\label{OEGC_MKN}
As for MS we can consider the angular position of the great circles. 
We find that for sufficiently large $\chi$ the equator is the great circle. Even when off-equator great circles develop
they do not asymptote to the poles for large $\mathcal{B}$ if $\chi \neq 0$. %As usual for large $\chi$ the geometry
%is KN-like.% while even for small positive $\chi$ the deformations aren't as extreme as for MS. 

We begin by looking for extremal values of the circumference.
Taking the derivative of $C$ we solve for its zeros to find:
\be
\frac{dC}{d \theta} = 0   \; \; \Longleftrightarrow \; \; \theta = \frac{\pi}{2} \; \; \mbox{or} \; \;  \cos^2 \! \theta = 1 - \frac{4 | \Lambda_o| \mathscr{R}^2}{ \mathcal{B}^2 R^2 (1-\chi^2)} \, . \label{GC}
\ee
There is always an extremum at the equator but there will also be two other extremum if the second expression has a 
solution. For a single root the equator has the largest circumference while if there are three roots the equator is a 
minimum and the other two are maxima . This is the same
pattern that we saw for MS. 

With $|\Lambda_o| \geq 1$, $\mathscr{R}/R \geq 1$ and $0 \leq \chi \leq 1$ it is straightforward to see 
that one must have $\mathcal{B} > 2$ in order for the equator to become a local minimum and the great circles to 
be located at $\theta \neq \frac{\pi}{2}$. The limiting case is MS and this is in agreement with our 
observations based on the Gaussian curvature\footnote{This of course is not surprising. Rewriting in terms of the 
circumference function
\be
\tilde{R} = \frac{(C'^2 - C C'') \sin \theta - C C' \cos \theta}{\pi^2 |\Lambda_o|^4 R^4 \sin^3 \! \theta} \, . 
\ee
Thus at the equator the Ricci scalar vanishes exactly where $C'' = 0$. }.

Next consider the great circles. For $\mathcal{B}>2$, $\frac{d}{d\mathcal{B}} (\cos^2 \! \theta) > 0$ and so if they diverge from
the equator, further increasing
$\mathcal{B}$ will always move the great circles further towards the poles. In the limit $\mathcal{B} = \pm \infty$ 
\be
\cos^2 \! \theta = 1 - \frac{\mathscr{R}^2 \sqrt{Q^4+4J^2} }{R^2 (1-\chi^2)}  \, . 
\ee
Thus they only asymptote to the poles for  MS. For $\chi \neq 0$ the angular position of the great circles asymptotes to 
a $\theta\subs{asymp} \neq \frac{\pi}{2}$. 

\subsubsection{Great circle versus equator} 
We can compare the ratio of the circumference of the equator $C\subs{eq}$ to that of the great circles $C\subs{gc}$. For 
MS we saw that $\lim_{B \rightarrow \pm \infty} {C\subs{eq}}/{C\subs{gc}} = 0$. However in the usual
way the MKN solutions are less extreme and the ratio goes to a finite value rather than zero. 

\begin{figure}
\begin{center}
\includegraphics[scale=0.2]{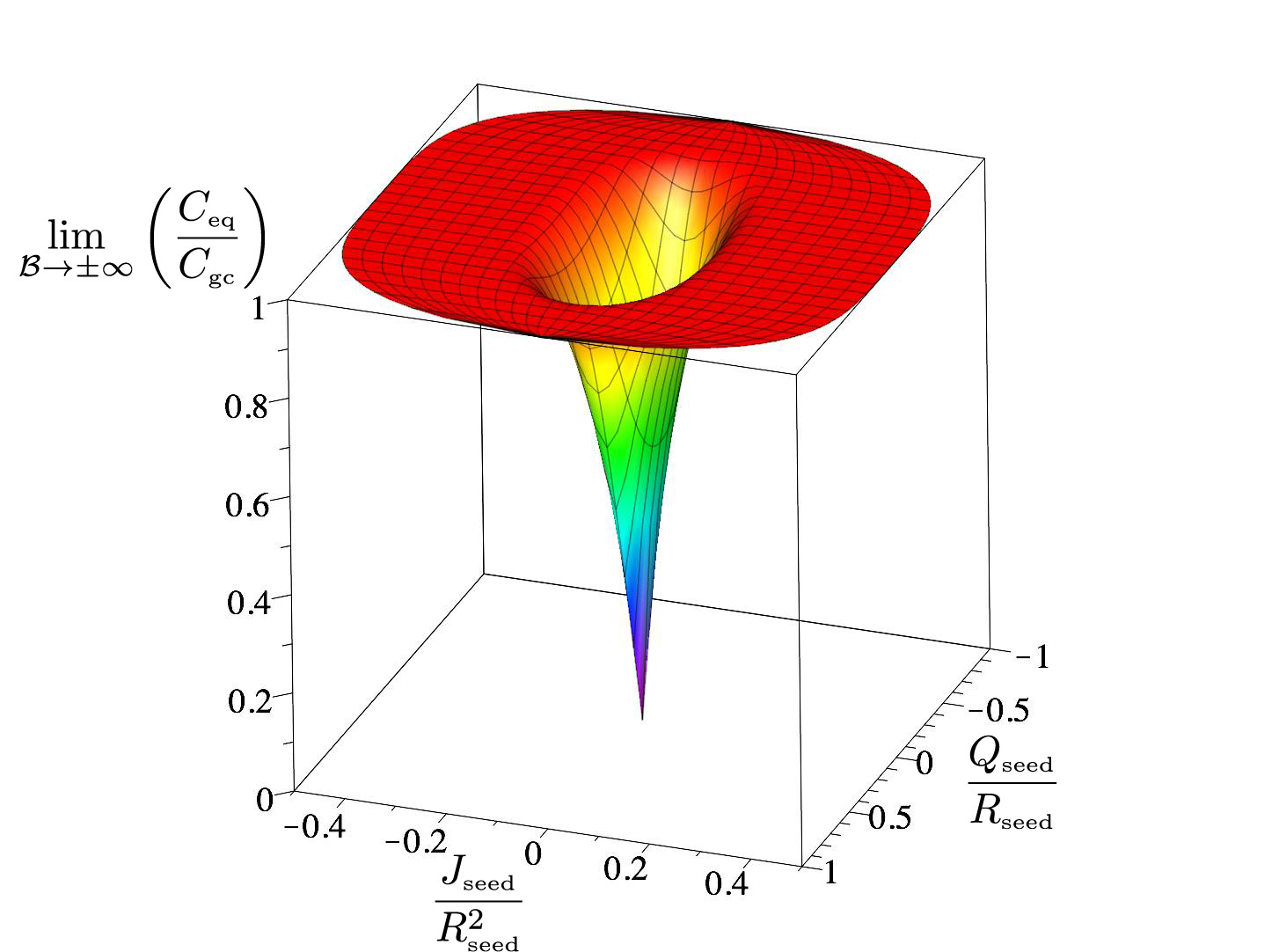}
\end{center}
\caption{Ratio of circumference of equator to that of the great circle for $\mathcal{B} \rightarrow \pm \infty$.}
\label{GCrat}
\end{figure}

In this case even the $\mathcal{B} \rightarrow \pm \infty$ limiting expression is very long and so we instead present
this result as a graph in Figure \ref{GCrat}. First looking at the extreme behaviours, for $\chi =0$ we recover the Melvin-
Schwarzschild limit while for large $\chi$ the ratio is unity (since the equator is the great circle). In-between the 
ratio is finite. 

\section{Conclusions}
\label{Sec:Conclusion}

In this paper we studied the geometry and physical properties of MKN horizons. In accord with known theorems we saw that their charge and angular momentum (both 
total and Komar) are bound by the areal radius in the same way as KN horizons. Also as expected extremal MKN horizons are geometrically identically to KN horizons
with the same area charge and angular momentum. 

Looking at the geometry of non-extremal horizons in more detail we saw that any non-zero $\chi$ imparts a degree of stiffness to the solutions. For small $\chi$
and large $\mathcal{B}$ MKN horizons still develop the characteristic MKN hour-glass shape however, in contrast to MS, they do not become arbitrarily long and thin. 
As $\mathcal{B} \rightarrow \infty$ they instead asymptote to well-defined finite endpoints. For $\chi \lesssim 1$ their shapes hew closely to KN horizons including
such key characteristics as regions of negative curvature at the poles. 

While many of these observations are qualitatively similar to \cite{Reiris:2013jaa} we worked with the degree of extremality $\chi$ rather than $J\subs{K}$.
At least for MKN horizons it is $\chi$ that provides more useful information. We expect that these observations can be extended to theorems
restricting the possible geometries of axisymmetric electrovac spacetimes however such work is left for future investigations.

\vspace{1cm}

\section*{Acknowledgements}
The authors would like to thank Marco Astorino for useful discussions on MKN spacetimes and in particular 
alerting us to several results in the literature of which we were not previously aware. 
IB and HK were respectively supported by NSERC grants 261429-2013 and 418537-2012.  
APL was supported by the CONICYT-FONDECYT/POSTDOCTORADO/3130674 project.

\appendix

\section{MKN functions on the horizon}
\label{AppMKNfunctions}

Exact, though complicated, forms for the various metric functions may be found in \cite{Gibbons:2013yq}. For our 
purposes we will only need to know to know these on the horizon and do not consider solutions with magnetic charge. 
Further, we are mainly interested in the forms of these quantities in terms of $(R,Q,J)$ rather than metric 
parameters $(m,q,a)$. The quantities below are evaluated directly from the expressions in Section \ref{FormSol} but may also be checked against explicit forms given in 
\cite{Gibbons:2013yq} by setting  setting $p=0$, $\Delta(r)=0 \Leftrightarrow m = \frac{1}{2} \left( r\subs{H}^2 + a^2 +q^2 \right)$ and then applying (\ref{ConvTrans}) to rewrite
in terms of the physical parameters of the seed solutions.  

First as noted in the text $\omega$ is constant on the horizon. That constant is:
\bea
\Omega\subs{H} &=&  X_o  + X_1 B + X_2 B^2
+ X_3B^3 + X_4 B^4 
\eea
where
\bea
X_o &=&
% \frac{a}{r\subs{H}^2 + a^2} 
\frac{2J}{R \mathscr{R}^2}\\ 
X_1 &=&%  -\frac{2qr\subs{H}}{r^2\subs{H}+a^2} 
 -\frac{2Q(R^2+Q^2)}{R \mathscr{R}^2}  \\
X_2 &=&
% -\frac{3aq^2}{2(r^2\subs{H}+a^2)}  
 -\frac{3JQ^2}{R \mathscr{R}^2}\\
X_3 &=&
% \frac{q}{2r\subs{H}} \left( q^2 \left( \frac{2 r\subs{H}^2 + a^2}{r\subs{H}^2 + a^2} \right) + (3r^2\subs{H}+a^2) \right) \\&=& 
\frac{Q}{2R \mathscr{R}^2} \left(3 R^4 + 5 Q^2 R^2 + 2 Q^4 + 4 J^2  \right)\\
X_4 &=& %\frac{a}{ 16r\subs{H}^2} \left( q^4 \left( \frac{3 r\subs{H}^2 + a^2}{r\subs{H}^2+a^2} \right)   + 2 q^2 (5r\subs{H}^2 + 2 a^2) + 2 (r\subs{H}^2+a^2)(3 r\subs{H}^2+a^2)  \right) \nonumber \\ & = & 
\frac{J}{8R\mathscr{R}^2} \left(6 R^4 + 10 Q^2 R^2 + 3 Q^4 + 8 J^2 \right) 
\eea
and for the quantities in terms of $(R,J,Q)$ we have applied (\ref{ConvTrans}). In the main text we will find it useful to 
rewrite
\bea
\Omega\subs{H} = \frac{8 J\subs{MKN} - 8QR^2 B + (2Q + BJ)(3\mathscr{R}^4- R^2 Q^2)  B^3}{4R \mathscr{R}^2}     \label{OmegaH}
\eea
with the help of (\ref{J_MKN}). 

The Coulomb potential  is also constant on the horizon and takes the form
%in terms of 
%$X_1, X_2, X_3$ and $X_4$:
%\bea
%\Phi_\xi = \frac{1}{2} \left( X_1  + 2  X_2  B + 3 X_3   B^2 + 4 B^3 \right) \, . 
%\eea
%That is 
\be
\Phi_\xi =  \left.  \xi^a A_a \right|\subs{horizon} = \frac{1}{2} \frac{d \Omega\subs{H}}{dB} \;  !
\ee
This somewhat surprising result follows from (\ref{E_MKN})-(\ref{Lambda}). 

Finally, from the  transformations of  Section \ref{FormSol}:
\be
f\subs{MKN} = \frac{f\subs{KN}}{| \Lambda |^2} \, , 
\ee
where $f\subs{KN}$ and $\Lambda$ are defined in the main text. 
%\be
%f\subs{KN} =  \frac{H \sin^2 \! \theta}{\Sigma}
%\ee
%where $H= (r^2 + a^2)^2 - \Delta a^2 \sin^2 \! \theta$, $\Sigma = r^2 + a^2 \cos^2 \! \theta$ and  $\Lambda = 1 + B \Phi  + \frac{1}{4} B^2 \mathcal{E}$. 
%This is quite a simple looking expression however if one expands and works only with real quantities, things becomes much more complicated, 
%even if we restrict our attention to the horizon. We find:
Then, at the horizon 
\be
\left. f\subs{MKN} \right|_{\mbox{\tiny{horizon}}} = \frac{(r\subs{H}^2 + a^2)^2 \sin^2 \! \theta}{G\subs{H} (\theta)}
\ee
where
%\be
%G\subs{H} (\theta) = G_0  + G_1  B + G_2  B^2  + G_3 B^3  + G_4 B^4 
%\ee
%for
%\be
%G_0 =  r\subs{H}^2 + a^2 \cos^2 \!  \theta \, , 
%\ee
%\be
%G_1 = 2qar\subs{H}\sin^2  \! \theta \, , 
%\ee
%\bea
%G_2  =  \frac{1}{2} \Big[ 3 q^2 \left(r\subs{H}^2 \cos^2 \theta + a^2 \right) +  (r\subs{H}^2 + a^2)^2 \sin^2  \!  \theta \Big] \, , 
%\eea
%
%\bea
%G_3 = \frac{aq}{2r\subs{H}} \Big[ q^2 ( 2 r\subs{H}^2 \cos^2  \! \theta  + a^2 \cos^2  \! \theta + a^2) + (r\subs{H}^2 + a^2)^2 (1+ \cos^2 \! \theta) \Big] 
%\eea
%and
%\bea
%G_4  =  \frac{1}{16}  & \Big[& q^4 \left((r\subs{H}^2 + a^2)^2 \cos^4 \! \theta + a^2 (3 r\subs{H}^2 + 2 a^2) \cos^2 \! \theta + a^4   \right)  \\
% & + & 2 q^2 (r\subs{H}^2 + a^2)^2 \left( - ( r\subs{H}^2 - a^2) \cos^4 \! \theta +  (r\subs{H}^2 + 2 a^2) \cos^2 \! \theta +  a^2 \right) \nonumber \\
% &+ &  (r\subs{H}^2 + a^2)^2  \left( (r\subs{H}^2 - a^2)^2 \cos^4 \! \theta - 2(r\subs{H}^4 - 2 a^2 r\subs{H}^2 - a^4)  \cos^2 \! \theta + (r\subs{H}^2 + a^2)^2 \right) \Big] \nonumber \, . 
%\eea
%
\be
G\subs{H} (\theta) = \frac{R^2}{\mathscr{R}^4} \left( G_0  + G_1  B + G_2  B^2  + G_3 B^3  + G_4 B^4 \right)
\ee
%for $\mathscr{R}^4 = (R^2+Q^2)^2 + 4 J^2$ and
with
\be
G_0 = (R^2+Q^2)^2 + 4J^2  \cos^2 \!  \theta \, , 
\ee
\be
G_1 = 4JQ(R^2+Q^2)\sin^2  \! \theta \, , 
\ee
\bea
G_2  &=&  \frac{1}{2} \Big[ Q^2(Q^2 R^2+2 R^4+12 J^2) +R^2 (R^4+4J^2) \sin^2 \! \theta \\
& &  + (3Q^4+5Q^2R^2+R^4)Q^2\cos^2 \! \theta \Big] \, , \nonumber
\eea
\bea
G_3 = JQ \Big[ (Q^2R^2+R^4+4J^2) + (2Q^4+3Q^2 R^2+R^4+4J^2)\cos^2 \! \theta \Big] 
\eea
and
\bea
G_4  & =  \frac{1}{16} &   \Big[ \Big( (R^4-Q^4-J^2)\cos^2 \! \theta + (R^4+Q^2R^2+4J^2) \Big)^2  \\
& & + 4J^2(Q^2+2R^2)^2 \cos^2 \! \theta  \Big]  \, .  \nonumber
\eea

\vspace{1cm}

\bibliographystyle{iopart-num}

\bibliography{../ArXiv/MKN_References}

\end{document}